\documentclass[aps,pra,twocolumn,showpacs,superscriptaddress,10pt,letterpaper]{revtex4-1}

\usepackage{graphicx}
\usepackage{dcolumn}
\usepackage{bm}
\usepackage{color}
\usepackage{amsmath}
\usepackage{mdframed}
\usepackage{enumitem}
\usepackage{subcaption}
\begin{document}

\title{On the Prevalence and Nature of Computational Instruction in Undergraduate Physics Programs across the United States}

\author{Marcos D. \surname{Caballero}}
\email[Corresponding Author: ]{caballero@pa.msu.edu}
\affiliation {Department of Physics and Astronomy \& CREATE for STEM Institute, Michigan State University, East Lansing, Michigan 48824}
\affiliation {Department of Physics \& Center for Computing in Science Education, University of Oslo, N-0316 Oslo, Norway}
\author{Laura Merner}
\affiliation{Statistical Research Center, American Institute of Physics, College Park, MD 20740}

\date{\today}

\begin{abstract}
A national survey of physics faculty was conducted to investigate the prevalence and nature of computational instruction in physics courses across the United States. 1246 faculty from 357 unique institutions responded to the survey. The results suggest that more faculty have some form of computational teaching experience than a decade ago, but it appears that this experience does not necessarily translate to computational instruction in undergraduate students' formal course work. Further, we find that formal programs in computational physics are absent from most departments. A majority of faculty do report using computation on homework and in projects, but few report using computation with interactive engagement methods in the classroom or on exams. Specific factors that underlie these results are the subject of future work, but we do find that there is a variation on the reported experience with computation and the highest degree that students can earn at the surveyed institutions.
\end{abstract}

\maketitle

\section{Introduction\label{sec:intro}}

The 21st century scientific world revolves around computation. For example, modeling core collapse supernovae, investigating pathways to chaos in electrical systems, and analyzing velocimetry data in mixing experiments, all require computational tools, methods, and practices.
Computing is part and parcel to many of the modern research efforts in physics. In fact, the development of new computational approaches and algorithms has helped support a number of important recent discoveries in physics including the Higgs' Boson \cite{aad2015combined} and the merger of black holes \cite{abbott2016observation}.
Including computation into physics courses is essential for the major to keep pace with current trends in research, industry, and, more broadly, an increasingly data-rich and model-driven society.
AIP reports that 75-90\% of  bachelors graduates are programming in their industrial positions while 50-60\% are performing some form of simulation and modeling \cite{mulvey2012physics}.
Recently, there have been calls to better represent computation in undergraduate coursework \cite{Behringer:2017up}.
More and better computational instruction has the potential to introduce students to physics as the discipline currently exists and to broaden career opportunities for graduating students.


A number of attempts have been made to integrate computation more fully into the undergraduate curriculum \cite{Dominguez:2015hd,Roos:2006vv,Timberlake:2004uz,Taylor:2006wq,Caballero:2012hu,mcintyre2008integrating,Chabay:2008jw,Cook:2008gh,Rebbi:2008gx,timberlake2008computation,Caballero:2014gn,Buffler:2008bf,irving2017p}. Many of the examples of this work were developed (and sustained) by faculty who have been engaged with computational physics as a core component of their professional research. To support incorporating computation more broadly, we must look beyond those faculty with research expertise in the area and to the rest of the faculty who are interested in and are likely supportive of computational instruction. Those faculty might be willing to make changes to their instruction, but, perhaps, they are constrained by time, energy, perceived expertise, or any combination of thereof.

This situation presents an opportunity for physics education researchers to support computational instruction by through institutional change research, the professional development of faculty, as well as the development of research-based instructional strategies and computational assessments. Because computational instruction is not yet widely used in physics, the field of computational physics education research is relatively wide-open. While we might draw on the more than 40 years of work in PER, specific research knowledge and instructional best practices are only beginning to be formed. What is presented in this paper is the initial landscape of computational instruction across the country.

As we consider how computation is included into coursework and how faculty choose to incorporate it into their courses, it becomes readily apparent that the state of computational instruction across the United States is a relative unknown. A survey of faculty across the country was conducted nearly a decade ago \cite{Fuller:2006wg,Chonacky:2008gq}, but how those initial findings have changed is important for the physics education community to understand. Furthermore, detailed information about faculty attitudes about and experiences with computation as well as any perceived barriers towards the greater use of computation are undocumented. In this research, we ask: what is the current state of computational instruction in physics departments across the country?

In this paper, we describe the results from a national survey of faculty that aimed to investigate the current state of computational instruction in undergraduate physics programs across the United States. Here, we report on the prevalence and nature of computational instruction in physics departments. That is, we have limited our presentation to reporting frequencies and to performing some cross-tabulations. For the sake of clarity and brevity, we have saved investigations of trends in faculty responses and any inferential analyses for future work. We should note that this survey was conducted in collaboration with the American Institute of Physics (AIP) by whom the second author (LM) is employed. AIP will be publishing its own report on this effort.

The paper is organized as follows: In Sec.~\ref{sec:motivation}, we present the motivation for the survey and situate it in the context of prior work. We discuss the design of the survey including its validation and the sampling process in Sec.~\ref{sec:design}. The results from the survey and their implications are discussed in Sec.~\ref{sec:results} and Sec.~\ref{sec:discussion} respectively. We conclude with important questions for future analyses in Sec.~\ref{sec:conclusions}.

\section{Motivation and Background\label{sec:motivation}}

Working to integrate computation into physics courses demands that we understand the nature of computational instruction including the current state of that instruction, what variations we observe in the implementation of computation, and what factors support or inhibit departments and faculty moving towards computational instruction. Chonacky and Winch conducted a survey of faculty over a decade ago \cite{Fuller:2006wg,Chonacky:2008gq}. The authors intended this survey to identify faculty early-adopters of computation, establish a detailed database of computational activities, confirm the importance attributed to computation in courses, and manifest how insular the beliefs that computation is important are within departments.

Responses to the survey were collected from 252 physics departments -- out of around 762 such departments at the time ($\sim$33\% response rate). The survey uncovered that most respondents ($\sim$80\%) agreed that ``{\it computed numerical approaches to learning physics principles ought to share the stage with analytic approaches}'' and, similarly, that most ($\sim$68\%) disagreed that ``{\it analytic approaches to learning physics principles are necessary and sufficient for educating physics student.}'' But, fewer that 20\% of the respondents reported including computation in their coursework. This work brought to light an interesting dilemma in the work of integrating computation into physics courses. Physics faculty recognize the importance of an educational experience that includes computation and, yet, the vast majority do not include computation in their courses.

Our current research aimed to expand on this work by designing a more thorough survey and by leveraging the expertise of AIP in its design and distribution. While the Chonacky and Winch survey provided some important information, it was quite limited in scope and depth -- containing roughly ten questions -- and was not validated in traditional sense of survey research \cite{fowler2013survey}. Additionally, we were interested to know if the state of computational instruction and the attitudes around it have changed in the last decade.

The work presented here represents the development of a survey in collaboration with AIP that aimed to investigate five areas:
\begin{enumerate}
  \item {\bf Current Uses} - Are faculty currently teaching computation in their physics courses? If so, in which courses and in what ways?
  \item {\bf Barriers and Successes} - What are the barriers to teaching computation at each institution? What have been the successes of teaching computation in the undergraduate curriculum?
  \item {\bf Faculty Attitudes} - How do faculty feel about computation? About teaching computation? What about their department?
  \item {\bf Future Plans} - What are faculty's future plans with regard to computational instruction in their department?
  \item {\bf Past Experience} - What are faculty's prior experiences with computation?
\end{enumerate}

As the survey questions span a substantial space, we limit our analysis in this paper to item 1. Future work will explore items 2-5 in more detail. By working with AIP, we were able to leverage the whole of AIP's experience in survey design and deployment. This ensured that we developed a rigorously-validated survey that lends itself to basic analytical approaches, which is what we present here. Inferential analysis that can offer explanations of these results presented here will be the subject of future work. In addition, we were able to distribute our survey easily to physics departments and faculty across the United States quite easily -- a definite challenge for Chonacky and Winch at the time \cite{Fuller:2006wg}.

\section{Survey Design and Distribution\label{sec:design}}

The design and distribution of the survey took place in 4 stages between fall 2015 and fall 2016: (1) development of scope, (2) construction of items, (3) validation with sample participants, and (4) distribution to the broader sample. In each of these stages, we worked to ensure the development of the survey fit well with the typical survey development process that AIP has followed for a number of years. This includes appropriate sampling processes and optimized timing for advertising, distribution, and collection based on AIP's own empirical evidence.

\paragraph{Development of Scope} We intended the results of this survey to be useful to a variety of stakeholders in the physics community. Thus, deciding what areas the survey should include and what areas it could leave out was not a decision to make in the absence of community. We convened a meeting of 24 stakeholders from across the United States with the mission of helping us to collectively develop and define the areas that the survey would probe. Industry professionals, physics faculty, curriculum designers, educational researchers, and recently-employed bachelors graduates met for a two-day working meeting to help the project team define the foci of the survey. We engaged participants with discussion across a variety of topics related to computational physics instruction in small groups -- all while taking field notes on the discussion. As we progressed over the two day period, nucleations of the aforementioned five areas became clear. In the final hours of the meeting, project staff presented these five areas to participants -- opening the floor to discussion and critiques. Participants generally agreed with the importance of the five areas over the (many) other possible topics discussed during the meeting. Some more senior participants offered advice in addressing items that might fall under each area, but no survey items were immediately developed during that meeting.

\paragraph{Construction of Items} With the five areas defined by stakeholders, the project team convened a two-day meeting at AIP headquarters to work to develop an initial draft of the questions. First, we worked to better define each area, and then to develop broad open-ended questions that could be classified under one or more areas. Questions such as: ``What are departments actually doing with computation in their undergraduate program?'' and ``If faculty value teaching computational physics to their students, why? That is, which aspects are valued?'' were developed to focus our discussion. This exercise helped to make sure that the project team were understanding each other's goals and that we could justify why to ask certain questions and not others. The resulting open-ended questions more narrowly constrained the scope of the survey and allowed us to work closely during a third, in-person meeting to draft individual survey items. Items were developed from these broad questions and by drawing on the tacit knowledge of AIP staff about the nature of appropriate survey questions. For example, the question ``What are departments actually doing with computation in their undergraduate program?'' became a single binary question (``Does your department, including efforts by individual faculty members, teach computational physics (see question below for examples) in its undergraduate curriculum?'') coupled to a series of binary questions (``If Yes, please select all the ways used to teach computational physics in your department:''). Through this process, we initially produced a set of survey items that was about 25\% longer than AIP suggested. These suggestions were based on their empirical evidence of survey fatigue. Several online meetings were held to cull and to combine redundant survey items to fit within AIP's guidelines. The resulting survey has a total of 187 items, which are subject to binary logic such that any one survey participant was likely to see less than 60 items.

\paragraph{Validation with Sample Participants}

AIP will typically distribute surveys to a small subset of the community that they are surveying. The rationale behind this is to validate the survey prior to sending it out more broadly. A sample survey was distributed to seven faculty who were representative of the faculty that might receive the survey. These validators were asked to take the survey and offer their feedback on scope and wording, as well as to articulate any confusion they had when completing the survey. Of these seven, we included three of the five faculty participants from the initial planning meeting. The result of these discussions with all seven faculty validators was that the survey items were mostly clear and interpretable, likely owing to AIP's experience in crafting such surveys. Minor critiques on wording were incorporated by AIP staff into the finalized version of the survey. After the validated version was produced, the survey was distributed online to a subset of the intended population to ensure that the survey itself collected appropriate data, that the logic worked properly, and that the resulting data collected online was interpretable.

\begin{figure}[t]
  \includegraphics[width=\columnwidth]{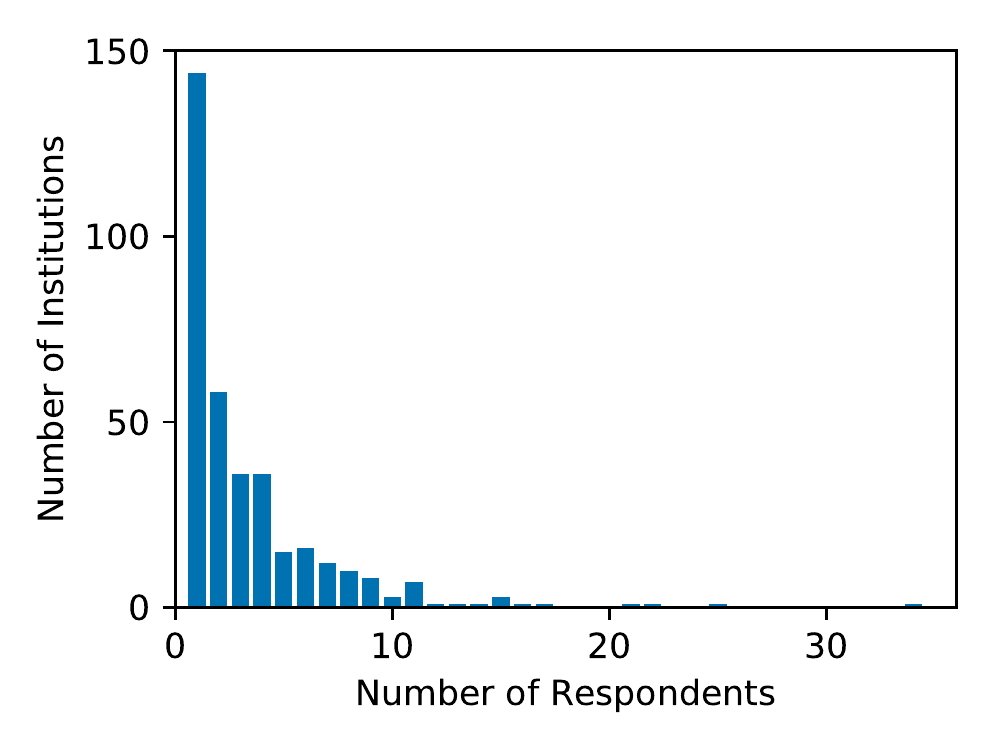}
  \caption{The number of institutions with a given number of respondents is shown. 40\% (144) of institutions had one faculty member respond to the survey while the remaining institutions (213) had more than one respondent including some institutions with more than 10 respondents and one with 35 respondents.}
  \label{fig:respondents}
\end{figure}

\paragraph{Distribution and Sampling Process}

The survey was distributed to a pseudo-random sampling of physics departments across the country. We ensured that a random sample of each category of institution for which AIP collects data was represented: Institutions offering terminal degrees at different levels: BS, MS, PhD; Research-intensive universities; Liberal arts and science Colleges; Two-year colleges; small, medium and large schools; as well as small, medium, and large graduating classes of physics majors. In the United States, there are around 750 institutions that offer bachelor's degrees in physics of which $\sim$66\% are predominantly undergraduate institutions; $\sim$7\% are Master's institutions; and $\sim$25\% are PhD granting universities. In addition, there are nearly 1500 two year colleges in the US. So within each of these categories our selection process was random, but we intended to sample from each category. A random sample of faculty within a selected institution were contacted to complete the survey. Ultimately, our data set contained responses from 1246 faculty at 357 unique institutions. In our sample, 139 institutions were two year colleges (9\% of total), 153 were undergraduate institutions (30\% of total), 18 were Master's institutions (33\% of total), and 47 were PhD institutions (25\% of total). Fig.~\ref{fig:respondents} shows the number of institutions with given number of respondents. 40\% of institutions (144) had only one faculty respond to the survey while the remaining 213 had more than one respondent including some departments with as many as 10 or more faculty responding to the survey.


\section{Results\label{sec:results}}

The data collected from these 1246 faculty was analyzed to investigate issues associated with item 1 (Sec.~\ref{sec:motivation}). Reporting these data is difficult given the variation of responses from faculty at a single intuition. That is, reporting results in terms of the percentage of respondents overweights the responses of larger departments or departments who had more faculty respond to the survey. In addition, within faculty in a given department, there was sometimes disagreement on the some of the most fundamental questions (i.e., ``Does your department offer a degree in computational physics?''). Hence, we have decided to present the data in two forms: (1) the fraction of departments with at least 50\% surveyed faculty responding positively and (2) the fraction of departments with at least 1 faculty member responding positively. Obviously, the fraction of departments reported in the second format is equal to or larger than the first format.

Faculty taking the survey were presented with the following inclusive definition of computation,

\begin{quote}
For the purposes of this survey, we are taking a broad view of computation, which includes a wide spectrum of examples such as: having students work with simulations and/or algorithms, giving students pieces of code to complete on their own, and/or advising students on undergraduate research projects where they write code from scratch.
\end{quote}

Defining computation for faculty in this way was done in an attempt to ensure that all faculty were working from the same definition when completing the survey. This approach was strongly suggested by the participants in the two-day workshop as each participant had their own definition of computation. By making this definition inclusive, we suggest that these data represent an upper-limit on the prevalence of computational instruction.

\begin{figure}[t]
  \includegraphics[width=\columnwidth]{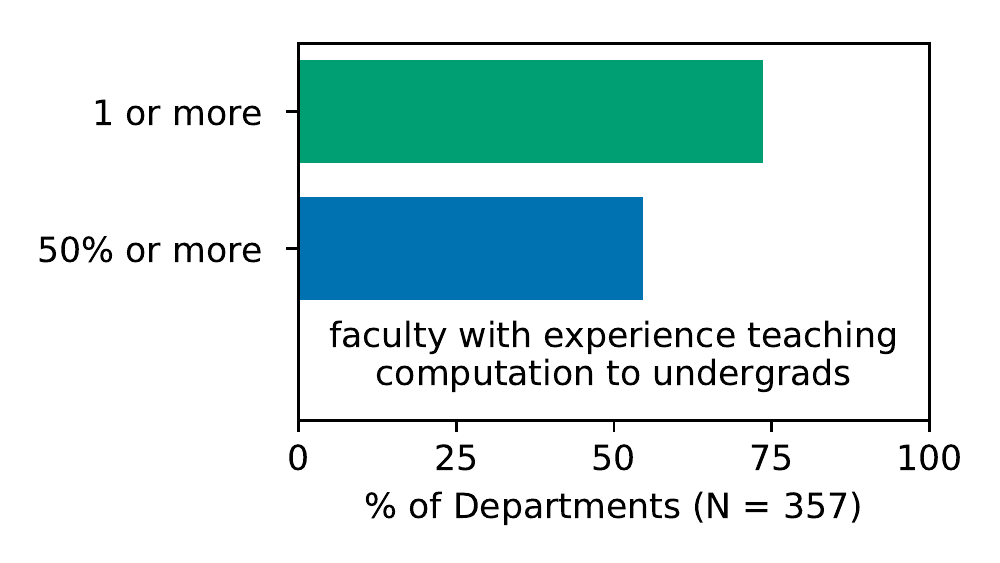}
  \caption{Out of the 357 unique departments represented in the survey, 195 (55\%) have at least 50\% of their faculty reporting that they have some experience teaching computation to undergraduate physics students (blue bar). 263 (74\%) departments have at least one faculty member reporting this experience (green bar).}
  \label{fig:teachComp}
\end{figure}

In Fig.~\ref{fig:teachComp}, we find that at least 50\% of surveyed faculty at 55\% of institutions ($N_{tot} = 357$) responded that they have experience teaching computation to undergraduate physics students (blue bar). However, 74\% of departments had at least one faculty member who responded that they have experience teaching computation to undergraduates (green bar). 

\begin{figure}[t]
    \centering
        \includegraphics[width=\linewidth]{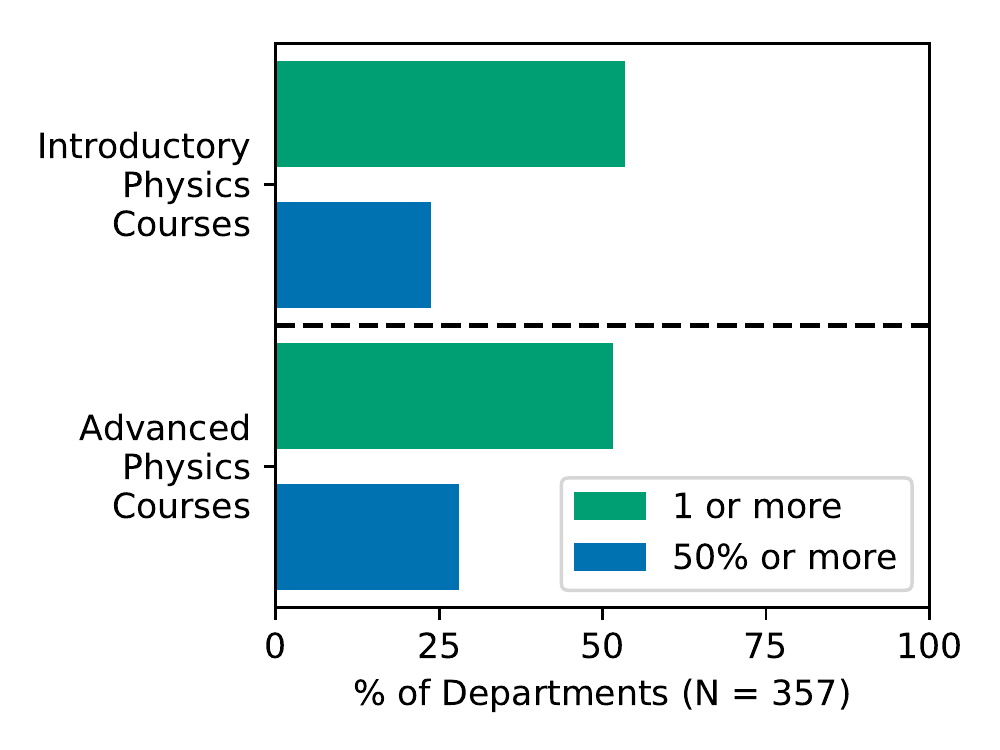}

    \caption{Out of the 357 unique departments, 100 (28\%) have at least 50\% of faculty reporting that they teach computation in an advanced-level physics course (upper half; blue bar). 184 (52\%) departments have at least one faculty member reporting that they teach computation in an introductory-level physics course (upper half; green bar). 85 (24\%) have at least 50\% of faculty reporting that they teach computation in an introductory-level physics course (lower half; blue bar). 191 (54\%) departments have at least one faculty member reporting that they teach computation in an introductory-level physics course (lower half; green bar).}
    \label{fig:introvadv}
\end{figure}

While a significant number of faculty report to have experience teaching computation to undergraduates (Fig.~\ref{fig:teachComp}), the proportion of faculty that report that their departments teach computation to undergraduates in formal course work is lower. For introductory-level courses, 24\% of departments have at least 50\% of surveyed faculty reporting that computation is taught in this courses (Fig.~\ref{fig:introvadv}. This fraction is similar to departments with more than 50\% of faculty reporting the computation is taught in some advanced-level courses (also Fig.~\ref{fig:introvadv}). In fact, a contingency table analysis of these data suggest that there is no association with the proportion of departments reported to teach computation and the level of the course ($\chi^2 = 1.43$, $p = 0.23$, $\nu = 1$). However, we find that the fraction of departments with at least one faculty member reporting that computation is taught in introductory or advanced courses is just above 50\% for both and similarly there is no association between the frequency of reporting and course level.

\begin{figure}[t]
    \centering
    \includegraphics[width=\linewidth]{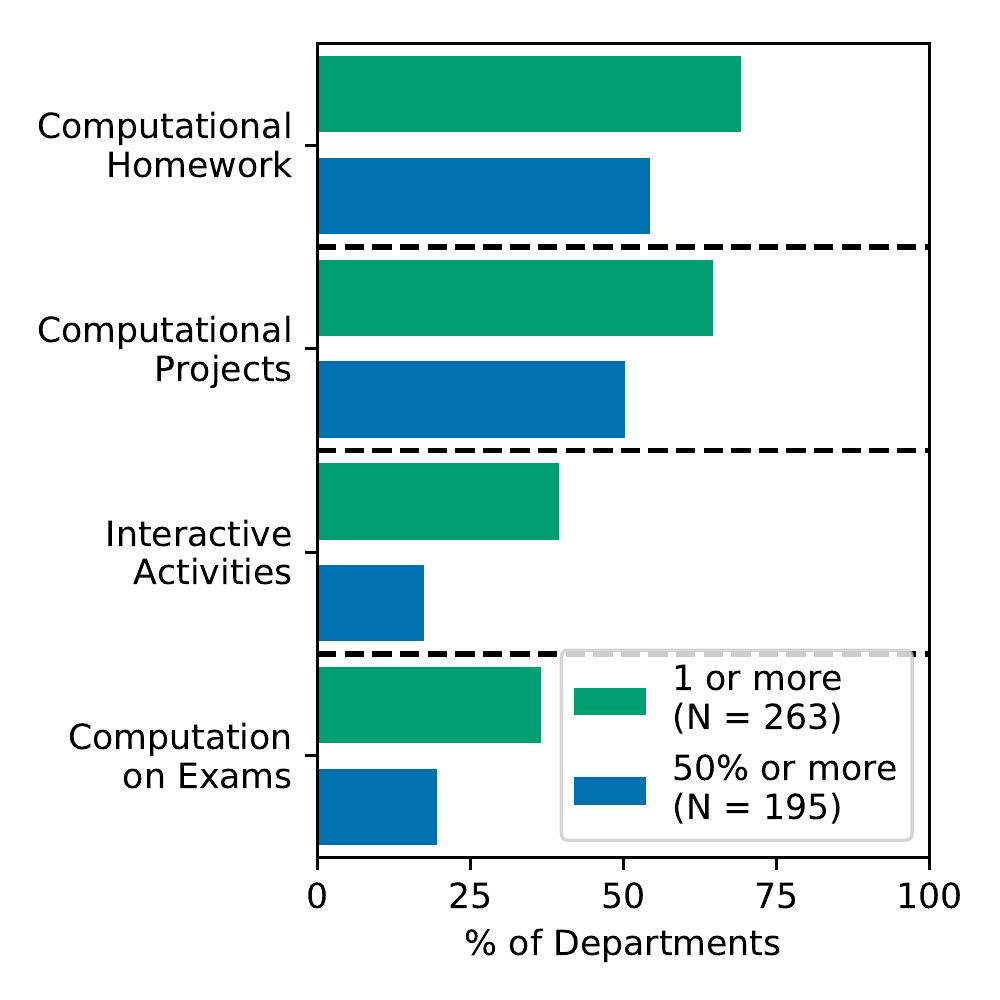}
    \caption{For the 195 departments that have at least 50\% of surveyed faculty reporting experience teaching computation in undergraduate physics, the percentage of departments with at least 50\% of surveyed faculty using computational homework problems, computational projects, interactive computational activities, and computational exam problems are shown (blue bars). For the 263 departments with at least 1 faculty member reporting this teaching experience, the percentage of departments with at least 1 surveyed faculty member reporting the use of these course materials is also shown (green bars).}
    \label{fig:methods}
\end{figure}

The specifics of the courses in which computation was taught were not directly captured. Hence, these courses reported in Fig.~\ref{fig:introvadv} might be any number of courses taught in the typical undergraduate curriculum, or, something altogether different. Open-response questions were asked, but few faculty chose to complete those questions making it difficult to infer trends in the broader data. However, specific questions about the nature of the tasks that students complete in these courses were asked of all faculty who responded that computation was taught in some course. Faculty were asked if homework, projects, interactive activities, and exams that made use of computation were used in these courses.

\begin{figure}[t]
    \centering
    \includegraphics[width=\linewidth]{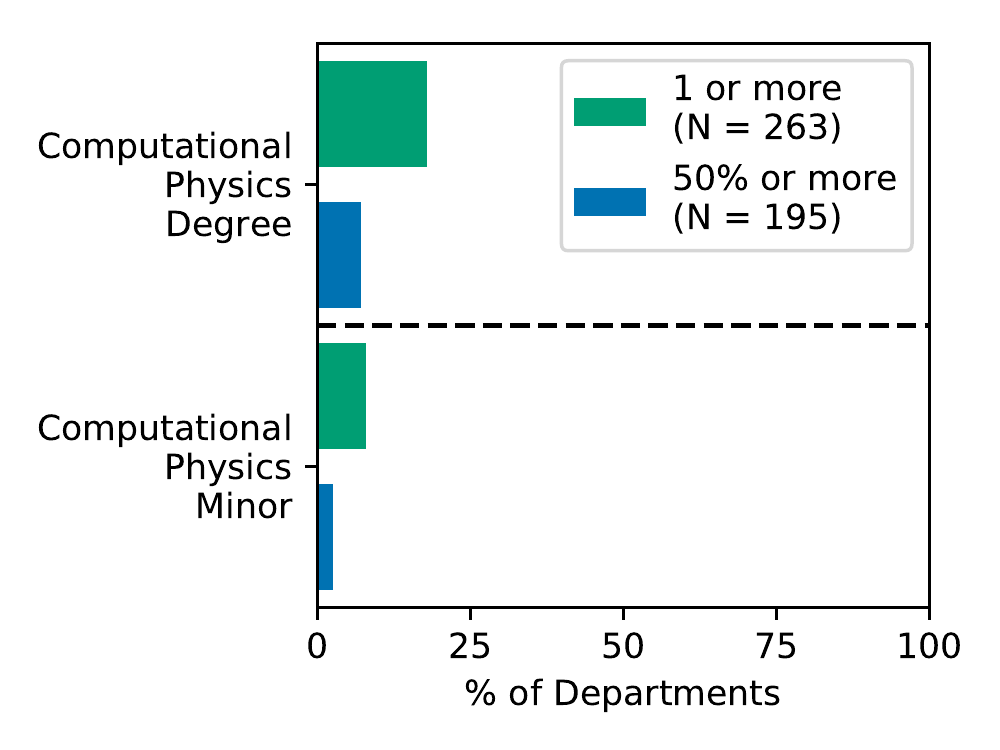}
    \caption{For the 195 departments that have at least 50\% of surveyed faculty reporting experience teaching computation in undergraduate physics, programs with at least 50\% of surveyed faculty reporting that they department offers bachelor's degrees in computational physics or minors in computational physics are shown (blue bars). The percentage of departments with at least 1 surveyed faculty member reporting that their department offers these formal progams is also shown (green bars).}
    \label{fig:formal}
\end{figure}

From Fig.~\ref{fig:methods}, we find that out of the 195 departments with at least 50\% of surveyed faculty reporting that computation is taught, slightly more than 50\% (106 departments) report that some form of computational homework is used.  This fraction is increased to 63\% when we instead use the number of departments with at least one faculty member reporting the use of computational homework. The number of departments in this analysis increases from 195 to 263, but the number of departments where one faculty member reports using computational homework increases disproportionately to 182. We find a similar fraction of departments (98 departments) that have at least 50\% of faculty reporting using computational projects. Again, this proportion grows as we include any department with one faculty member reporting that they use computational projects. Again, there are 263 such departments with 170 having at least one faculty member reporting the use of computational projects.

For classroom activities that make use of some form of interactive engagement \cite{Meltzer:2012eg}, we find fewer faculty report using such activities with computation than homework or projects (Fig.~\ref{fig:methods}). Less than 25\% of departments have at least 50\% of faculty report using interactive engagement activities with computation. This proportion does increase to just below 40\% when we consider a single faculty member's response. Similarly, the reporting of assessment of computation through the use of examinations is low compared to homework and projects. Less than 25\% departments have 50\% or more of their faculty report assessing computation on exams. This fraction increases to 35\%, as the inset indicates, when considering the responses of a single faculty member.

Some departments offer formal programs to support computationally-interested students including majors and minors in computational physics. The percentage of departments offering majors and minors (Fig.~\ref{fig:formal}) are shown for the 195 departments with at least 50\% of surveyed faculty reporting experience teaching computation. The percentages of such departments are quite low, 7\% and 3\% respectively. These percentages increase slightly for both majors (18\%) and minors (8\%) when considering the report of a single faculty member (263 departments).

There are a number of factors that could be mediating the results presented in Figs. \ref{fig:teachComp}-\ref{fig:formal}. The type of institution including its resources and focus could limit or enhance the use of computation in different courses. Faculty have a variety of backgrounds and opinions about computation that might further mediate these outcomes. Given the scope of this paper, a detailed exploration of the mediating factors using inferential analysis will be the subject of future work.

However, we can observe a clear variation in the reporting of experience with computational instruction by an institution's highest degree (Fig.~\ref{fig:bydegree}). Here, we observe that faculty teaching at institutions offering Associates degrees report the least experience with teaching computation while those at Bachelors, Masters, and Doctoral institutions report more experience. That is, the fraction of departments with more than 50\% of faculty reporting experience with computation varies with institution type. A contingency table analysis ($\chi^2 = 27.38$; $p \ll 0.05$; $\nu = 3$) shows an association with the prevalence of computational teaching experience and an institution's highest available degree. However, pairwise $\chi^2$ comparisons using a Bonferroni correction ($p<0.0125$) demonstrate that this effect is only due to a lower fraction of two-year college faculty reporting they have computational teaching experience. In this data, there is no statistically significant association between percentage of departments with at least 50\% of respondents reporting computational teaching experience and different classes of institutions that grant 4 year physics degrees.

\section{Discussion\label{sec:discussion}}

From the analysis presented in Sec.~\ref{sec:results}, we find wide variation among faculty responses within individual departments about factors that one might assume most members of the faculty within that department would be aware (e.g., Does the department offer a degree in computational physics?). To deal with this variation, we have reported an upper-limit to the prevalence of computational instruction by using any single faculty member's report as representing the whole department (green bars in Figs.~\ref{fig:teachComp}--\ref{fig:formal}). We have also offered a conservative estimate of the same by counting results only when more than 50\% of respondents from a department respond in the affirmative. Additionally, we believe it is likely that those faculty who completed the survey were those who were more likely to be computational users -- simply because the nature of the survey was one that would be of interest to such faculty. Thus, for the purposes of discussing the current state of computational instruction in departments, we will use this conservative estimate to both address the observed variation and expected bias.

\begin{figure}[t]
  \includegraphics[width=0.95\columnwidth]{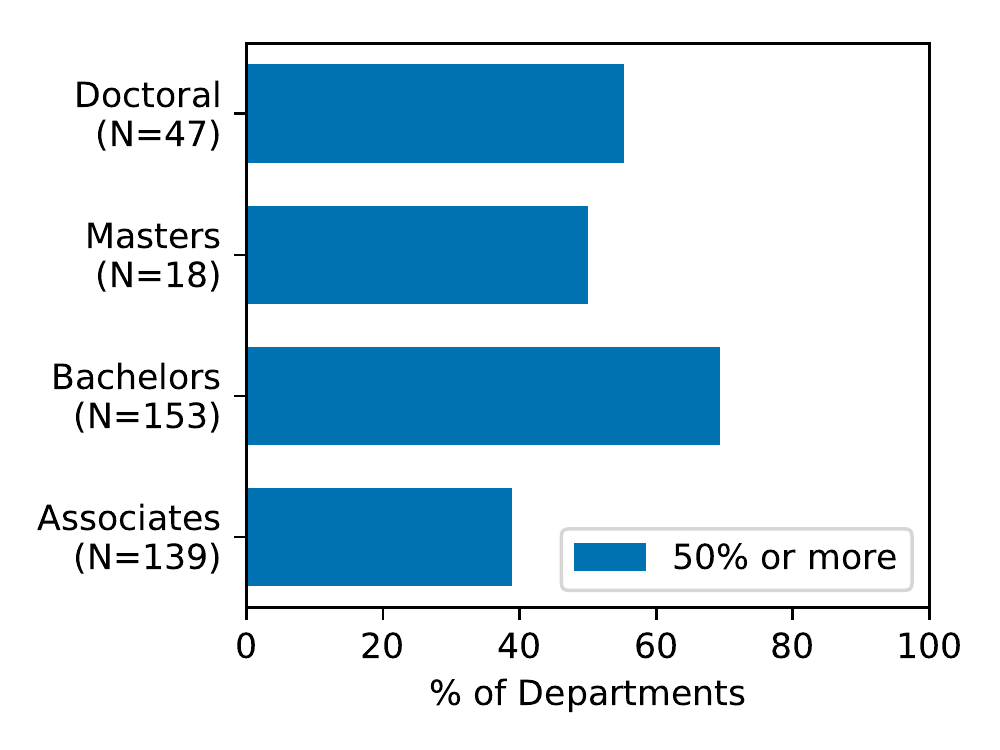}
  \caption{The percentage of departments with at least 50\% of surveyed faculty at an institution reporting that they have experience teaching computation varies by the highest degree that students can earn in physics.}
  \label{fig:bydegree}
\end{figure}

We find that a majority of faculty respondents report some experience teaching computation to undergraduate students and that a majority of departments have at least 50\% of faculty reporting having such experience (Fig.~\ref{fig:teachComp}). This demonstrates a growth in faculty expertise over the last decade based on the survey conducted by Chonacky and Winch \cite{Fuller:2006wg,Chonacky:2008gq}. However, this increase in reported experience does not appear to translate over entirely to teaching computation in formal courses (Fig.~\ref{fig:introvadv}). Around 25\% of departments report teaching computation in some introductory or some advanced course. What is interesting to note is that the fraction of departments reporting teaching computation in either introductory or advanced courses does not differ significantly. Hence, there is an opportunity for physics education researchers to help translate faculty experience into practice at all levels of computational instruction.

With regard to the nature of the instruction, we find that most departments have faculty using computation on homework assignments and in project-based formats. Such activities are typically designed for students to complete outside of class on their own time with support from faculty coming in different forms based on the faculty member's availability, schedule, and teaching practices. Less than 25\% of departments have faculty reporting that they use interactive activities for computational instruction in class or present students with computational problems on exams. The first of these activities could, in principle, borrow from substantial literature on active-learning in physics \cite{Meltzer:2012eg}, but the specific techniques for teaching computation interactively and the necessary tools to do so (e.g., clicker questions and tutorial activities) have not been broadly developed and disseminated. There is a unique opportunity to develop and test interactive methods for computational instruction at a variety of course levels, which are grounded in both learning theory and best practices for instruction. Here, physics education research can offer significant support through research studies on how students approach computational work and what challenges they might face \cite{Caballero:2012hu,Obsniuk:2015ks}. Moreover, the low fraction of departments with faculty who report using computational questions on exams is suggestive of an additional challenge -- the nature of computational assessment. Assessing how students have learned computation is not something that readily translates over to standard examination practices in physics. This kind of summative assessment could be quite difficult to conduct, especially at scale. As such, there is an additional opportunity to develop assessments and assessment practices that can investigate what students have learned after an experience with computation in a physics course.

Finally, the specific features that inhibit or support computational instruction are likely to stem from departmental and faculty factors. Indeed, it is likely the case that important factors are combinations of resources, background, experience, and personal opinions -- all of which are interrelated in some way. While we have presented one of these factors (Fig.~\ref{fig:bydegree}), a detailed analysis that unpacks and discusses these features will be presented in future work. For the purposes of this work, we do find, as an example, that there might be an opportunity for professional development with two-year college faculty. That being said, how to support all college faculty who are interested in incorporating computation is an open question and an area that physics education researchers might support through by curriculum and professional development activities.

\section{Conclusion and Future Work\label{sec:conclusions}}

We have reported on a survey aimed at determining the prevalence and nature of computational instruction in physics courses across the United States. In conducting this survey, we received responses from 1246 faculty at 357 unique institutions. This work demonstrates that the needle has moved slightly with regard to instruction in computation, but that there is a significant opportunity for the PER community to support the development of materials, teaching practices, and assessment strategies across all levels of computational physics instruction. Furthermore, there is a dire need to perform research in these environments at a variety of scales. With regard to this survey, additional work is needed to examine the underlying factors that are suggestive of whether and how faculty choose to teach computation in physics courses. The barriers that faculty perceive, the resources that might support their move toward integrating computation, and the attitudes that faculty hold about computation are all data collected by the survey. Future work will investigate how these factors are indicative of efforts to teach computation in physics courses. Models of these data should provide suggestions for faculty and departments looking to support the integrating computation into their physics courses.

\begin{acknowledgments}
The authors would like to thank Norman Chonacky and Robert Hilborn for their work to organize the stakeholder meeting and the development of survey areas. The authors would also like to thank members of PERL@MSU for their helpful comments on drafts of this paper. This work was supported by the National Science Foundation's Division of Undergraduate Education (DUE-1431776 and DUE-1432363).
\end{acknowledgments}

\bibliography{refs}

\begin{thebibliography}{22}
\expandafter\ifx\csname natexlab\endcsname\relax\def\natexlab#1{#1}\fi
\expandafter\ifx\csname bibnamefont\endcsname\relax
  \def\bibnamefont#1{#1}\fi
\expandafter\ifx\csname bibfnamefont\endcsname\relax
  \def\bibfnamefont#1{#1}\fi
\expandafter\ifx\csname citenamefont\endcsname\relax
  \def\citenamefont#1{#1}\fi
\expandafter\ifx\csname url\endcsname\relax
  \def\url#1{\texttt{#1}}\fi
\expandafter\ifx\csname urlprefix\endcsname\relax\def\urlprefix{URL }\fi
\providecommand{\bibinfo}[2]{#2}
\providecommand{\eprint}[2][]{\url{#2}}

\bibitem[{\citenamefont{Aad et~al.}(2015)\citenamefont{Aad, Abbott, Abdallah,
  Abdinov, Aben, Abolins, AbouZeid, Abramowicz, Abreu, Abreu
  et~al.}}]{aad2015combined}
\bibinfo{author}{\bibfnamefont{G.}~\bibnamefont{Aad}},
  \bibinfo{author}{\bibfnamefont{B.}~\bibnamefont{Abbott}},
  \bibinfo{author}{\bibfnamefont{J.}~\bibnamefont{Abdallah}},
  \bibinfo{author}{\bibfnamefont{O.}~\bibnamefont{Abdinov}},
  \bibinfo{author}{\bibfnamefont{R.}~\bibnamefont{Aben}},
  \bibinfo{author}{\bibfnamefont{M.}~\bibnamefont{Abolins}},
  \bibinfo{author}{\bibfnamefont{O.}~\bibnamefont{AbouZeid}},
  \bibinfo{author}{\bibfnamefont{H.}~\bibnamefont{Abramowicz}},
  \bibinfo{author}{\bibfnamefont{H.}~\bibnamefont{Abreu}},
  \bibinfo{author}{\bibfnamefont{R.}~\bibnamefont{Abreu}},
  \bibnamefont{et~al.}, \emph{\bibinfo{title}{Combined measurement of the higgs
  boson mass in p p collisions at s= 7 and 8 tev with the atlas and cms
  experiments}}, \bibinfo{journal}{Physical review letters}
  \textbf{\bibinfo{volume}{114}}, \bibinfo{pages}{191803}
  (\bibinfo{year}{2015}).

\bibitem[{\citenamefont{Abbott et~al.}(2016)\citenamefont{Abbott, Abbott,
  Abbott, Abernathy, Acernese, Ackley, Adams, Adams, Addesso, Adhikari
  et~al.}}]{abbott2016observation}
\bibinfo{author}{\bibfnamefont{B.~P.} \bibnamefont{Abbott}},
  \bibinfo{author}{\bibfnamefont{R.}~\bibnamefont{Abbott}},
  \bibinfo{author}{\bibfnamefont{T.}~\bibnamefont{Abbott}},
  \bibinfo{author}{\bibfnamefont{M.}~\bibnamefont{Abernathy}},
  \bibinfo{author}{\bibfnamefont{F.}~\bibnamefont{Acernese}},
  \bibinfo{author}{\bibfnamefont{K.}~\bibnamefont{Ackley}},
  \bibinfo{author}{\bibfnamefont{C.}~\bibnamefont{Adams}},
  \bibinfo{author}{\bibfnamefont{T.}~\bibnamefont{Adams}},
  \bibinfo{author}{\bibfnamefont{P.}~\bibnamefont{Addesso}},
  \bibinfo{author}{\bibfnamefont{R.}~\bibnamefont{Adhikari}},
  \bibnamefont{et~al.}, \emph{\bibinfo{title}{Observation of gravitational
  waves from a binary black hole merger}}, \bibinfo{journal}{Physical review
  letters} \textbf{\bibinfo{volume}{116}}, \bibinfo{pages}{061102}
  (\bibinfo{year}{2016}).

\bibitem[{\citenamefont{Mulvey and Pold}(2017)}]{mulvey2012physics}
\bibinfo{author}{\bibfnamefont{P.~J.} \bibnamefont{Mulvey}} \bibnamefont{and}
  \bibinfo{author}{\bibfnamefont{J.}~\bibnamefont{Pold}},
  \emph{\bibinfo{title}{Physics bachelors: Initial employment}}
  (\bibinfo{year}{2017}).

\bibitem[{\citenamefont{Behringer}(2017)}]{Behringer:2017up}
\bibinfo{author}{\bibfnamefont{E.}~\bibnamefont{Behringer}},
  \emph{\bibinfo{title}{{AAPT Recommendations for Computational Physics in
  Undergraduate Physics Curricula}}}, \bibinfo{journal}{Bulletin of the
  American Physical Society} \textbf{\bibinfo{volume}{Volume 62, Number 6}}
  (\bibinfo{year}{2017}).

\bibitem[{\citenamefont{Dominguez and Huff}(2015)}]{Dominguez:2015hd}
\bibinfo{author}{\bibfnamefont{R.}~\bibnamefont{Dominguez}} \bibnamefont{and}
  \bibinfo{author}{\bibfnamefont{B.}~\bibnamefont{Huff}},
  \emph{\bibinfo{title}{{The role of computational physics in the liberal arts
  curriculum}}}, \bibinfo{journal}{Journal of Physics: Conference Series}
  \textbf{\bibinfo{volume}{640}}, \bibinfo{pages}{012061}
  (\bibinfo{year}{2015}).

\bibitem[{\citenamefont{Roos}(2006)}]{Roos:2006vv}
\bibinfo{author}{\bibfnamefont{K.~R.} \bibnamefont{Roos}},
  \emph{\bibinfo{title}{{An incremental approach to computational physics
  education}}}, \bibinfo{journal}{Computing in science {\&} engineering}
  \textbf{\bibinfo{volume}{8}}, \bibinfo{pages}{44} (\bibinfo{year}{2006}).

\bibitem[{\citenamefont{Timberlake}(2004)}]{Timberlake:2004uz}
\bibinfo{author}{\bibfnamefont{T.}~\bibnamefont{Timberlake}},
  \emph{\bibinfo{title}{{A computational approach to teaching conservative
  chaos}}}, \bibinfo{journal}{American Journal of Physics}
  \textbf{\bibinfo{volume}{72}}, \bibinfo{pages}{1002} (\bibinfo{year}{2004}).

\bibitem[{\citenamefont{Taylor and King}(2006)}]{Taylor:2006wq}
\bibinfo{author}{\bibfnamefont{J.~R.} \bibnamefont{Taylor}} \bibnamefont{and}
  \bibinfo{author}{\bibfnamefont{B.~A.} \bibnamefont{King}},
  \emph{\bibinfo{title}{{Using computational methods to reinvigorate an
  undergraduate physics curriculum}}}, \bibinfo{journal}{Computing in science
  {\&} engineering} \textbf{\bibinfo{volume}{8}}, \bibinfo{pages}{38}
  (\bibinfo{year}{2006}).

\bibitem[{\citenamefont{Caballero et~al.}(2012)\citenamefont{Caballero,
  Kohlmyer, and Schatz}}]{Caballero:2012hu}
\bibinfo{author}{\bibfnamefont{M.~D.} \bibnamefont{Caballero}},
  \bibinfo{author}{\bibfnamefont{M.~A.} \bibnamefont{Kohlmyer}},
  \bibnamefont{and} \bibinfo{author}{\bibfnamefont{M.~F.}
  \bibnamefont{Schatz}}, \emph{\bibinfo{title}{{Implementing and assessing
  computational modeling in introductory mechanics}}},
  \bibinfo{journal}{Physical Review Special Topics - Physics Education
  Research} \textbf{\bibinfo{volume}{8}}, \bibinfo{pages}{020106}
  (\bibinfo{year}{2012}).

\bibitem[{\citenamefont{McIntyre et~al.}(2008)\citenamefont{McIntyre, Tate, and
  Manogue}}]{mcintyre2008integrating}
\bibinfo{author}{\bibfnamefont{D.~H.} \bibnamefont{McIntyre}},
  \bibinfo{author}{\bibfnamefont{J.}~\bibnamefont{Tate}}, \bibnamefont{and}
  \bibinfo{author}{\bibfnamefont{C.~A.} \bibnamefont{Manogue}},
  \emph{\bibinfo{title}{{Integrating computational activities into the
  upper-level Paradigms in Physics curriculum at Oregon State University}}},
  \bibinfo{journal}{American Journal of Physics} \textbf{\bibinfo{volume}{76}},
  \bibinfo{pages}{340} (\bibinfo{year}{2008}).

\bibitem[{\citenamefont{Chabay and Sherwood}(2008)}]{Chabay:2008jw}
\bibinfo{author}{\bibfnamefont{R.}~\bibnamefont{Chabay}} \bibnamefont{and}
  \bibinfo{author}{\bibfnamefont{B.}~\bibnamefont{Sherwood}},
  \emph{\bibinfo{title}{{Computational physics in the introductory
  calculus-based course}}}, \bibinfo{journal}{American Journal of Physics}
  \textbf{\bibinfo{volume}{76}}, \bibinfo{pages}{307} (\bibinfo{year}{2008}).

\bibitem[{\citenamefont{Cook}(2008)}]{Cook:2008gh}
\bibinfo{author}{\bibfnamefont{D.~M.} \bibnamefont{Cook}},
  \emph{\bibinfo{title}{{Computation in undergraduate physics: The Lawrence
  approach}}}, \bibinfo{journal}{American Journal of Physics}
  \textbf{\bibinfo{volume}{76}}, \bibinfo{pages}{321} (\bibinfo{year}{2008}).

\bibitem[{\citenamefont{Rebbi}(2008)}]{Rebbi:2008gx}
\bibinfo{author}{\bibfnamefont{C.}~\bibnamefont{Rebbi}},
  \emph{\bibinfo{title}{{A project-oriented course in computational physics:
  Algorithms, parallel computing, and graphics}}}, \bibinfo{journal}{American
  Journal of Physics} \textbf{\bibinfo{volume}{76}}, \bibinfo{pages}{314}
  (\bibinfo{year}{2008}).

\bibitem[{\citenamefont{Timberlake and
  Hasbun}(2008)}]{timberlake2008computation}
\bibinfo{author}{\bibfnamefont{T.}~\bibnamefont{Timberlake}} \bibnamefont{and}
  \bibinfo{author}{\bibfnamefont{J.~E.} \bibnamefont{Hasbun}},
  \emph{\bibinfo{title}{{Computation in classical mechanics}}},
  \bibinfo{journal}{American Journal of Physics} \textbf{\bibinfo{volume}{76}},
  \bibinfo{pages}{334} (\bibinfo{year}{2008}).

\bibitem[{\citenamefont{Caballero and Pollock}(2014)}]{Caballero:2014gn}
\bibinfo{author}{\bibfnamefont{M.~D.} \bibnamefont{Caballero}}
  \bibnamefont{and} \bibinfo{author}{\bibfnamefont{S.~J.}
  \bibnamefont{Pollock}}, \emph{\bibinfo{title}{{A model for incorporating
  computation without changing the course: An example from middle-division
  classical mechanics}}}, \bibinfo{journal}{American Journal of Physics}
  \textbf{\bibinfo{volume}{82}}, \bibinfo{pages}{231} (\bibinfo{year}{2014}).

\bibitem[{\citenamefont{Buffler et~al.}(2008)\citenamefont{Buffler, Pillay,
  Lubben, and Fearick}}]{Buffler:2008bf}
\bibinfo{author}{\bibfnamefont{A.}~\bibnamefont{Buffler}},
  \bibinfo{author}{\bibfnamefont{S.}~\bibnamefont{Pillay}},
  \bibinfo{author}{\bibfnamefont{F.}~\bibnamefont{Lubben}}, \bibnamefont{and}
  \bibinfo{author}{\bibfnamefont{R.}~\bibnamefont{Fearick}},
  \emph{\bibinfo{title}{{A model-based view of physics for computational
  activities in the introductory physics course}}}, \bibinfo{journal}{American
  Journal of Physics} \textbf{\bibinfo{volume}{76}}, \bibinfo{pages}{431}
  (\bibinfo{year}{2008}).

\bibitem[{\citenamefont{Irving et~al.}(2017)\citenamefont{Irving, Obsniuk, and
  Caballero}}]{irving2017p}
\bibinfo{author}{\bibfnamefont{P.~W.} \bibnamefont{Irving}},
  \bibinfo{author}{\bibfnamefont{M.}~\bibnamefont{Obsniuk}}, \bibnamefont{and}
  \bibinfo{author}{\bibfnamefont{M.}~\bibnamefont{Caballero}},
  \emph{\bibinfo{title}{P-cubed: a practice focused learning environment}},
  \bibinfo{journal}{European Journal of Physics}  (\bibinfo{year}{2017}).

\bibitem[{\citenamefont{Fuller}(2006)}]{Fuller:2006wg}
\bibinfo{author}{\bibfnamefont{R.}~\bibnamefont{Fuller}},
  \emph{\bibinfo{title}{{Numerical computations in US undergraduate physics
  courses}}}, \bibinfo{journal}{Computing in science {\&} engineering}
  \textbf{\bibinfo{volume}{8}}, \bibinfo{pages}{16} (\bibinfo{year}{2006}).

\bibitem[{\citenamefont{Chonacky and Winch}(2008)}]{Chonacky:2008gq}
\bibinfo{author}{\bibfnamefont{N.}~\bibnamefont{Chonacky}} \bibnamefont{and}
  \bibinfo{author}{\bibfnamefont{D.}~\bibnamefont{Winch}},
  \emph{\bibinfo{title}{{Integrating computation into the undergraduate
  curriculum: A vision and guidelines for future developments}}},
  \bibinfo{journal}{American Journal of Physics} \textbf{\bibinfo{volume}{76}},
  \bibinfo{pages}{327} (\bibinfo{year}{2008}).

\bibitem[{\citenamefont{Fowler~Jr}(2013)}]{fowler2013survey}
\bibinfo{author}{\bibfnamefont{F.~J.} \bibnamefont{Fowler~Jr}},
  \emph{\bibinfo{title}{Survey research methods}} (\bibinfo{publisher}{Sage
  publications}, \bibinfo{year}{2013}).

\bibitem[{\citenamefont{Meltzer and Thornton}(2012)}]{Meltzer:2012eg}
\bibinfo{author}{\bibfnamefont{D.~E.} \bibnamefont{Meltzer}} \bibnamefont{and}
  \bibinfo{author}{\bibfnamefont{R.~K.} \bibnamefont{Thornton}},
  \emph{\bibinfo{title}{{Resource Letter ALIP--1: Active-Learning Instruction
  in Physics}}}, \bibinfo{journal}{American Journal of Physics}
  \textbf{\bibinfo{volume}{80}}, \bibinfo{pages}{478} (\bibinfo{year}{2012}).

\bibitem[{\citenamefont{Obsniuk et~al.}(2015)\citenamefont{Obsniuk, Irving, and
  Caballero}}]{Obsniuk:2015ks}
\bibinfo{author}{\bibfnamefont{M.~J.} \bibnamefont{Obsniuk}},
  \bibinfo{author}{\bibfnamefont{P.~W.} \bibnamefont{Irving}},
  \bibnamefont{and} \bibinfo{author}{\bibfnamefont{M.~D.}
  \bibnamefont{Caballero}}, in \emph{\bibinfo{booktitle}{2015 Physics Education
  Research Conference}} (\bibinfo{publisher}{American Association of Physics
  Teachers}, \bibinfo{year}{2015}), pp. \bibinfo{pages}{239--242}.

\end{thebibliography}
\bibliographystyle{apsper}

\end{document}